\documentclass[12pt]{article}
\usepackage{pic03}
\usepackage{graphicx}

\begin{document}

\title{\bf PRECISION DRIFT CHAMBERS FOR THE ATLAS MUON SPECTROMETER}
\author{
S.Horvat, O.Kortner, H.Kroha, A.Manz, S.Mohrdieck, V.Zhuravlov        \\
{\em Max-Planck-Institut f\"{u}r Physik, F\"{o}hringer Ring 6, D-80805 M\"{u}nchen, Germany}}
\maketitle

\baselineskip=14.5pt
\begin{abstract}
ATLAS is a detector under construction to explore the physics at the Large Hadron Collider at CERN. It has a muon spectrometer with an excellent momentum resolution of 3-10\%, provided by three layers of precision monitored-drift-tube chambers in a toroidal magnetic field. A single drift tube measures a track point with a mean resolution close to 100~$\mu$m, even at the expected high n and $\gamma$ background rates. The tubes are positioned within the chamber with an  accuracy of 20~$\mu$m, achieved by elaborate construction and assembly monitoring procedures.
\end{abstract}

\baselineskip=17pt

\section{Introduction}
The ATLAS detector at the Large Hadron Collider (LHC) will explore the Higgs sector of the standard model as well as new physics at energy scales of 1-10 ~TeV. The ATLAS muon spectrometer provides a momentum resolution of 3-10\% for momenta between 10 and 1000~GeV/c~\cite{muon_tdr}. This is achieved by the precise reconstruction of the muon trajectories with three stations of  monitored-drift-tube (MDT) chambers in a typically 0.4~T toroidal magnetic field over distances of 6-10~m. The MDT chambers consist of 3 or 4 layers of cylindrical aluminum drift tubes of 30~mm diameter with only 0.4~mm wall thickness, on each side of an aluminum support frame. The tubes contain 50~$\mu$m diameter gold-plated W-Re sense wires.  They are operated with Ar:CO$_{2}$ (93:7) gas mixture at 3~bar pressure and a gas gain of 2$\cdot$10$^{4}$. Each drift tube measures a track point with an average resolution of 100~$\mu$m. With this resolution, the required tracking precision is achieved provided the positions of the sense wires within the chamber are known with an accuracy of 20~$\mu$m (rms).

\section{Construction of Monitored Drift Tube Chambers} 
\begin{figure}[htbp]
\hspace{0cm}\begin{minipage}[c]{7.5cm} 
\centerline{\hbox{\includegraphics[width=6.8cm]{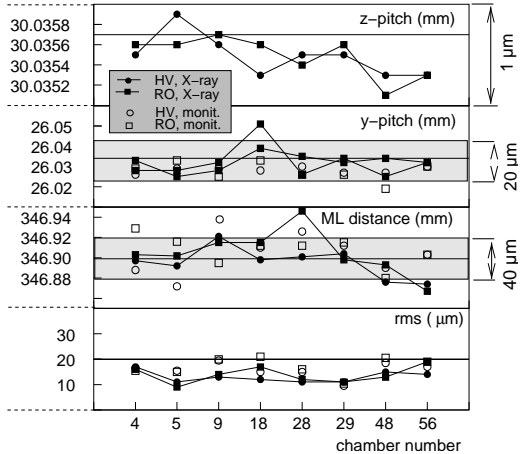}}}
\end{minipage}\begin{minipage}[c]{7.5cm} 
\caption{\it X-ray measurements (full symbols) of the wire grid parameters at the high-voltage (HV) and the readout (RO) end of the MDT chambers, in comparison with the monitoring during the construction (open symbols). The last row shows the standard deviations of the wire positions with respect to the nominal wire grid as obtained from the X-ray measurement and from the optical measurements.  The horizontal lines represent the nominal values. 
 \label{xtomo}}\end{minipage}
\end{figure}

The MDT chambers are assembled on a granite table with precise mechanical tools. The tube layers are glued consecutively to the support frame. The mechanical accuracy of the chambers is monitored with several monitoring techniques. As an example, we describe the chamber assembly at the Max Planck Institute f\"ur Physik~\cite{assembly}. The sense wires are centered in the individual tubes with an accuracy of 7~$\mu$m (rms) as verified by X-ray measurement of each tube. The 72 tubes of a layer are then positioned with an accuracy of 5~$\mu$m on the assembly table. The positioning is verified with mechanical feeler gauges. The support frame is positioned on the table with an accuracy of 5~$\mu$m and glued to the tube layer on the table. The relative positioning of the glued tube layers is monitored with optical RASNIK~\cite{rasnik} sensors. The gravitational deformations of the frame are measured by another RASNIK system and compensated during the gluing by computer-controlled pneumatic actuators. The monitoring techniques allow for the verification of the required wire-positioning accuracy of 20~$\mu$m for all chambers. In addition, for about 15\% of the chambers, the wire positions are measuremed with an X-ray scanning device at CERN~\cite{xtomo}. The geometrical parameters of the wire grid, the wire pitch in the direction parallel ($z$) and vertical ($y$) to the assembly table and the distance between the triple layers (ML) are reproducible and in good agreement with the nominal values (see Fig.~\ref{xtomo}).    

\section{MDT Chamber Performance under LHC Operating Conditions}

\begin{figure}[htbp]
\begin{minipage}[r]{11cm} 
\centerline{\hbox{\includegraphics[width=10cm]{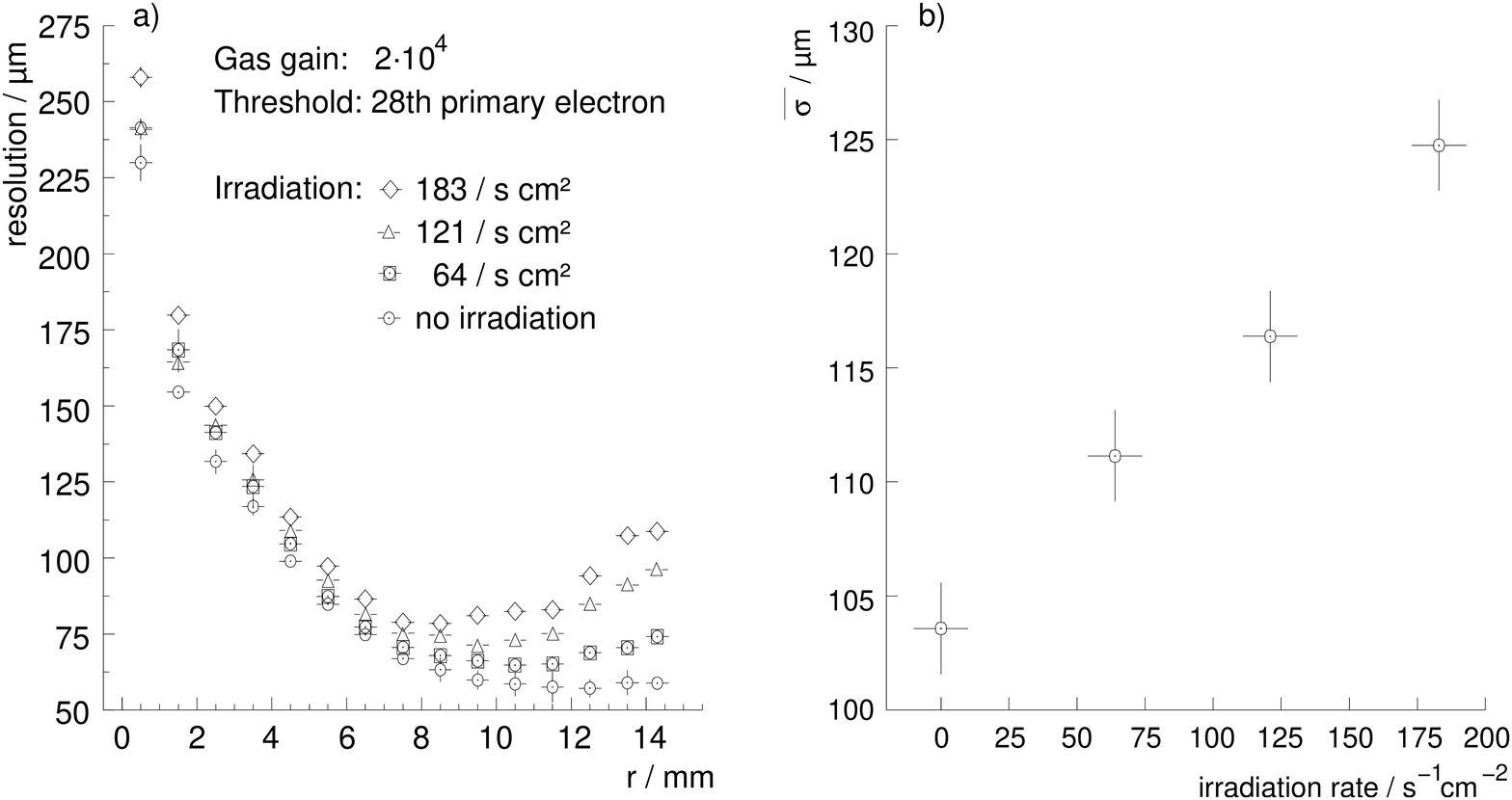}}}
\end{minipage}\begin{minipage}[c]{4cm} 
 \caption{\it a) Spatial resolution of  a single drift tube as a function of the impact radius r for different $\gamma$ irradiation rates. b) Average spatial resolution $\bar{\sigma}$ of a drift tube as a function of the $\gamma$ irradiation rate.
\label{x5}}\end{minipage}
\end{figure}

At the LHC, the chambers will be exposed to a high neutron and $\gamma$ background with counting rates of up to 100~Hz/cm$^{2}$. The performance of an MDT chamber at LHC background count rates was tested under nominal operating conditions in the $\gamma$ irradiation facility at CERN with a $^{137}$Cs source and a 100~GeV muon beam. The spatial resolution of a single drift tube is computed as $\sigma (r)=\sqrt{Var(r-r_{S})}$ where $r$ is the impact radius measured in the drift tube and $r_{S}$ the position extrapolated from a silicon strip tracking detector, used as external reference. Fig.~\ref{x5} shows the
resolution of a drift tube as a function of $r$ and the average tube resolution by different $\gamma$ irradiation rates. The resolution is degraded with increasing irradiation due to the increasing effect of space charge fluctuations on the drift time (see~\cite{kortner}). However, even at the highest expected rate of 100~Hz/cm$^{2}$, a high average spatial resolution of 114~$\mu$m - only 10~$\mu$m worse than without irradiation - is achieved. The single tube resolution is expected to improve with the final read-out electronics.

\end{document}